# Single electron capacitance spectroscopy of vertical quantum dots using a single electron transistor.

M. Koltonyuk, D. Berman[a] , N. B. Zhitenev[b] and R. C. Ashoori
*Department of Physics, Massachusetts Institute of Technology, Cambridge, MA 02139*

L. N. Pfeiffer and K. W. West
*Lucent Technologies Bell Laboratories, Murray Hill, NJ 07974*

We have incorporated an aluminum single electron transistor (SET) directly on top of a vertical quantum dot, enabling the use of the SET as an electrometer that is extremely responsive to the motion of charge into and out of the dot. Charge induced on the SET central island from single electron additions to the dot modulates the SET output, and we describe two methods for demodulation that permit quantitative extraction of the quantum dot capacitance signal. The two methods produce closely similar results for the determined single electron capacitance peaks.

The single electron transistor (SET)[1,2,3] is a highly charge-sensitive device, capable of detecting charges far less than that of one electron. This remarkable property allows an SET to be used as an extremely responsive electrometer, making it a useful tool in experiments where very high charge sensitivity is required[4,5,6]. In this Letter, we describe a novel method for using an SET as a charge sensor to study single electron addition spectra of quantum dots.

Detection of single electrons has proven to be a powerful tool for understanding the physics of quantum dots and other small structures[7,8]. Single-electron capacitance spectroscopy[9] performed with conventional FETs has permitted measurement of a few thousand individual electron additions in a single dot, starting from the first one[10]. In place of a FET, we now position a SET directly on top of the dot, thereby creating a possibility for much greater charge sensitivity. This arrangement is well suited for detecting *any* motion of charge under material's surface; it is flexible enough to allow study of charge motion in defects and impurities as well as a number of quantum dot structures.

The SET consists of a small central island coupled through two tunnel junctions to the source and drain electrodes (Fig. 1a). The operation of the SET is based on the Coulomb blockade principle, and the current passing through it can be varied by adjusting the mean charge on the central island using a gate[11]. Keeping the bias voltage $V_{S\text{-}D}$ fixed and varying the bottom electrode voltage $V_B$ leads to a periodic change in current through the SET (Fig. 1d). Movement of small charges near the central island of the SET also alters the current through the SET[6].

In our experiment, we study a type of vertical quantum dot[12,13] (Fig. 1c) confined between two capacitor plates with the SET directly on top of it. The dot is close enough to the bottom plates to allow tunneling of electrons between the dot and the plate. The SET can sensitively detect these subsurface charge motions.

We fabricated the SET on the surface of an MBE grown GaAs/AlGaAs heterostructure[9] (the parameters of the growth are discussed in Ref. 7) with a quantum well 80 nm below the surface, using the standard double-angle shadow evaporation method to produce the tunnel junctions[2,6]. Then, the central island was used as a mask for wet etching down to the AlGaAs blocking barrier (Fig. 1c), so as to confine the quantum dot to be directly under our SET[9]. By adjusting the voltage on the SET leads with respect to the bottom electrode we control the tunneling of electrons into the quantum dot.

We performed our experiments with the sample at 300 mK and in zero magnetic field so that the SET was superconducting for enhanced sensitivity[14]. The source-drain bias is adjusted for optimal gain[15]. Then, we apply a DC voltage to the bottom electrode together with a small superimposed AC excitation (20μV rms, 17Hz). The SET current is measured using a current lock-in amplifier. The DC voltage is then slowly varied, and the current through the SET is recorded resulting in a signal shown in Fig. 2a. The frequency of the measurement is sufficiently low so that all signals from the SET are in phase with the AC excitation, and the observed in-phase signal is directly proportional to the product of charge induced on the SET central island and $dI_{SET} \big/ dQ_{\substack{central \\ island}}$ .

The SET signal exhibits modulation in both frequency and amplitude as evident from our data (Fig. 2a). The fast oscillations are due to a relatively large capacitance linking directly the bottom electrode and the central island. Sweeping the DC voltage induces single electron additions on the SET central island and causes the amplitude of the AC current through the SET to oscillate (Fig. 1d).

The oscillation amplitude changes noticeably over a period of a few oscillations due to electron addition to the quantum dot (Fig. 2a). In general, the charge response, $dQ_S$, induced on the SET central island by the AC excitation on the bottom electrode (acting as a gate), $dV_B$, can be expressed as:

$$dQ_S = \left[ C_{B-S} + \frac{C_{D-S}}{C_{D-S} + C_{B-D}} \left( C_{B-D} + \frac{dQ_D}{dV_B} \right) \right] dV_B \,. \tag{1}$$

Here, $Q_D$ is the charge on the dot, $Q_S$ is the charge on the SET central island, and $V_B$ is the voltage on the bottom electrode ($C_{B-S}$ and all other capacitances are defined in the diagram in Fig. 1b). The change in current through the SET, $dI_S$, due to a changing $V_B$ is proportional to the above expression, excluding factors discussed below.

Electrons are added to the quantum dot at particular values of $V_B$, defining a “single electron addition spectrum”[13]. Between these values, and $dQ_D / dV_B$ is zero in the above expression. When single electron tunneling does occur, the response of the SET changes drastically because at these points $dQ_D/dV_B$ becomes large. In fact, $dQ_D/dV_B$ behaves as a derivative of a Fermi function, with its peak approaching infinity at zero temperature.

At gate voltages for which a single electron is added to the dot, the effective gate capacitance is greatly increased due to the contribution from $dQ_D/dV_B$. This creates an additional "loading" capacitance on the SET central island, and the width of the Coulomb blockade region of the SET shrinks, and so does the SET gain. Further, as the SET optimal source-drain biasing condition depends on the central island capacitance, fluctuations in dot's potential cause the transistor to move away from conditions for maximum gain thus providing another reason for diminishment of the amplitude of the SET current oscillations. Finally, electron tunneling events in the dot also change the frequency of the SET oscillations. An electron addition causes additional charge to be induced on the SET central island and thereby increases the oscillation frequency.

To extract the quantum dot capacitance signal from our data, we can exploit either frequency (inverse spacing of oscillations as a function of $V_B$) or amplitude variation of the SET gain. The frequency of Coulomb blockade oscillations is directly proportional to the charge induced on the SET central island as a result of scanning the potential of the bottom electrode only, and does not depend on the shunting capacitance or any other experimental parameters. Therefore, we can directly extract the quantum dot capacitance peaks by measuring how the frequency changes when an electron tunnels into the dot, *independent* of any other experimental variables, such as temperature, biasing of the SET, etc. For this reason we consider this method of extracting the peaks superior to the one based on examination of the amplitude modulation. We account for the envelope modulation by scaling all data peaks to unity, so that only frequency modulation remains. A frequency modulated sine wave represents this data rather well, so that taking an inverse sine of the modified data reveals the phase change in oscillations due to electrons entering the dot, while the ensuing differentiation produces the desired capacitance peak. The results for two arbitrarily selected electrons entering the dot successively are shown in Fig. 2c.

Analysis of amplitude modulation of the envelope of the sine wave provides another method for determination of the capacitance. Using a simulation program[16,17], we are able to model the dependence of the gain of the SET should on shunt capacitance arising from to $dQ_D / dV_B$. We ran the program for several different values of that shunt capacitance while keeping all other SET operational parameters fixed. The program calculates SET gain. Thus, we construct theoretically the relationship between the SET gain and extra shunt capacitance (Fig. 2b). Subsequently, we apply the reverse transformation relating a drop in gain to an additional shunt capacitance, which we previously have obtained theoretically. Thus, we derive a capacitance peak corresponding to the addition of that electron. A precise determination of the peak amplitude is difficult using this method. The extracted peak height depends critically on the values used in the simulation program for temperature, lead and shunt capacitances. Nonetheless, the determined peak shape is robust to variations in these data since in the range of our operational parameters the relationship between the gain and the shunt capacitance is nearly linear (Fig. 2b). The capacitance peaks for the same two electrons as mentioned in the previous paragraph are shown in Fig. 2c (re-scaled to fit).

In summary, we have demonstrated a novel technique allowing a SET to be used as a charge sensor for objects buried within a semiconductor and developed two independent methods for interpreting the data. This method holds significant potential for applications the in other situations such studying of charge motion in defects and impurities, since it allows observing electrons move in a solid with extraordinary sensitivity.

This work was supported by the Office of Naval Research, the National Science Foundation DMR, David and Lucille Packard Foundation, and Joint Services Electronics program.

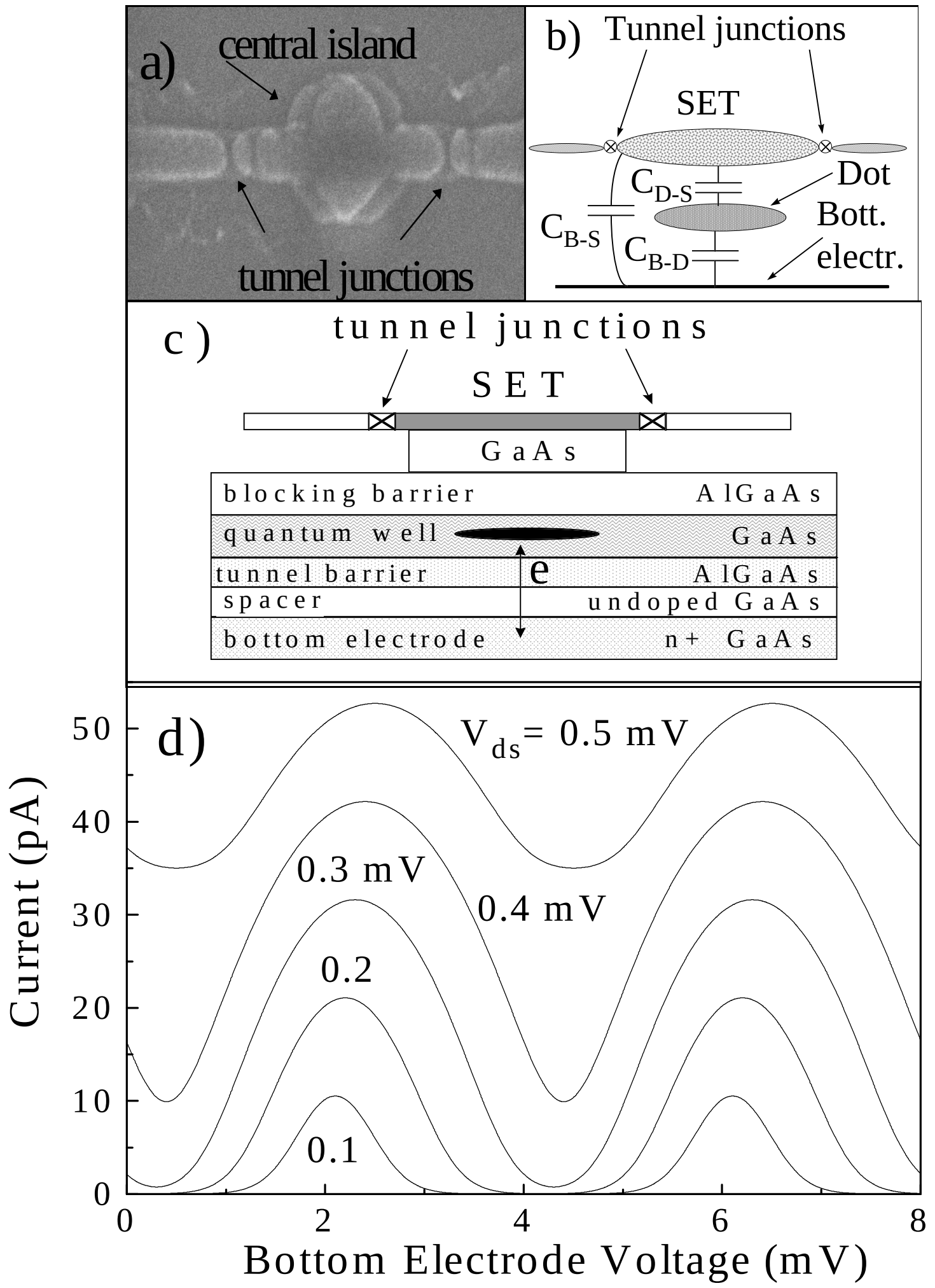

**Fig. 1.** (a) Scanning electron microscope photograph of the SET. The central island of the SET serves as an etch mask for producing a quantum dot which exists underneath it. (b) Schematic diagram for our sample. The quantum dot lies between the bottom electrode and the SET central island. (c) Diagram of the sample produced on an MBE grown wafer. (d) Dependence of the current through the SET on the gate voltage for several different drain-source biases.

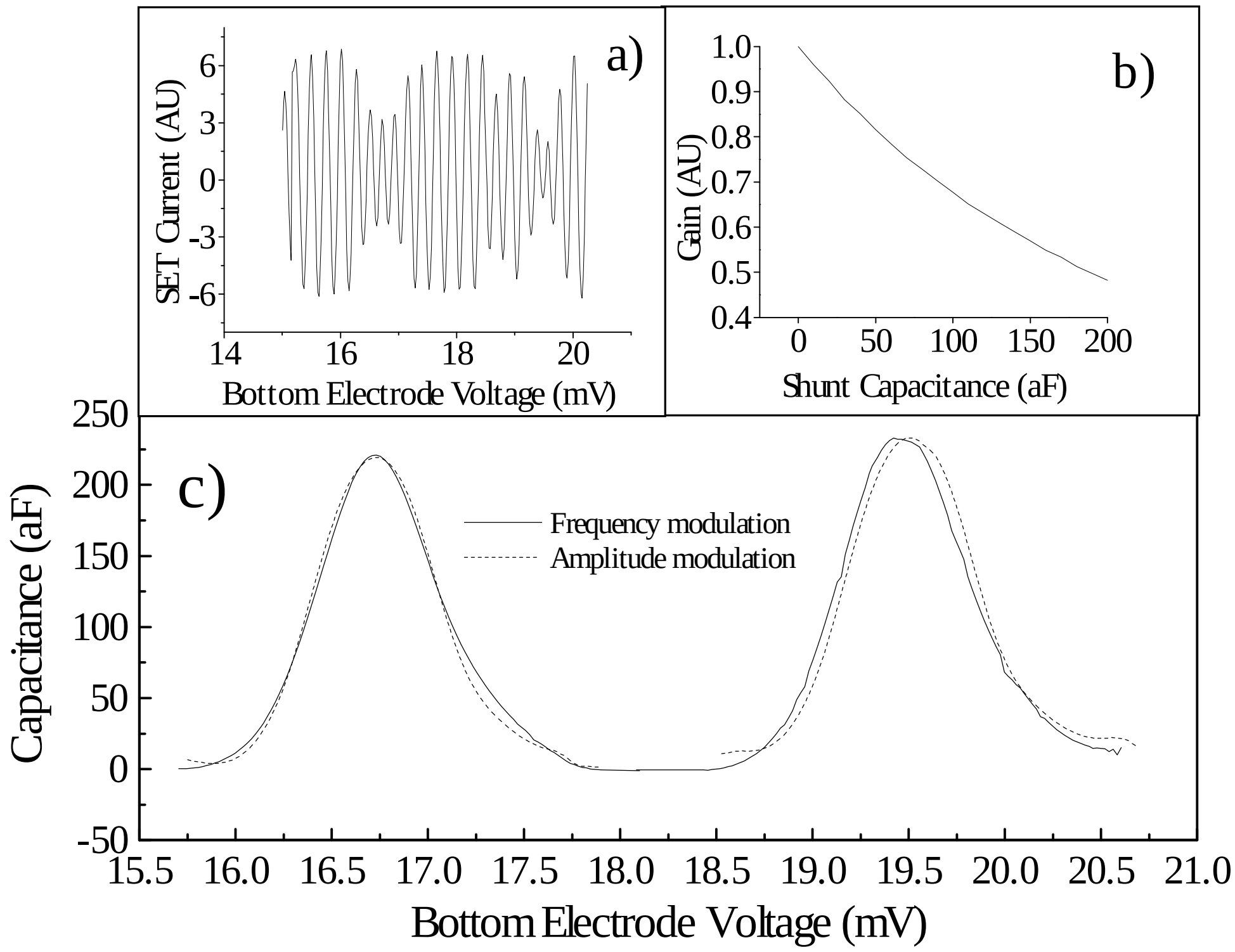

**Fig. 2.** (a) AC current amplitude through the SET as a function of voltage applied to the bottom electrode. Each oscillation corresponds to one electron addition on the SET central island. (b) Calculated dependence of the SET gain on additional shunt capacitance arising from electron tunneling into the quantum dot. (c) Two neighboring single electron additions detected by SET. The data extracted from frequency and amplitude demodulation of the measured SET current signal (the peak extracted from amplitude modulation is given in arbitrary vertical units).